# The Old and New Meanings of Cloud 'Belt' and 'Zone':

## A Study of Jovian and Saturnian Atmospheric Banding

## Based on Hubble Space Telescope Images

Anthony Mallama

Galileo Observatory, 14012 Lancaster Lane, Bowie, MD, 20715 USA

anthony.mallama@gmail.com

2014 December 23

**Abstract**

The brightness of cloud bands on Jupiter and Saturn as a function of latitude is reported. Bright Jovian bands near the equator are located in regions of anti-cyclonic circulation of the atmosphere. By contrast, bright equatorial bands on Saturn are associated with cyclonic motion. Modern definitions of the cloud band terms 'zone' and 'belt' are distinguished from their old meanings.

## 1. Introduction

Alternating bright and dark cloud bands encircle each of the giant planets. The clouds are associated with atmospheric circulation and the vertical motions of gases. This paper describes the empirical relationship between Jovian and Saturnian banding and atmospheric circulation. Brightness information was derived from Wide Field Planetary Camera images from the Hubble Space Telescope. The Saturnian results are fully described here while the Jovian results are excerpted from another manuscript.

Section 2 establishes the latitude limits of cyclonic and anti-cyclonic atmospheric circulation on Saturn. The methods of selecting HST images and of measuring brightness are described in section 3. The association between circulation and brightness is reported in section 4. In section 5, the circulation-brightness relationship for Saturn is compared with that for Jupiter.

This paper avoids using the terms 'zone' and 'belt' to distinguish between bright and dark bands, respectively. Section 6 explains that their historical definitions, which relate to brightness, no longer apply. In the modern literature, the terms 'zone' and 'belt' distinguish between anti-cyclonic and cyclonic circulation.

## 2. Atmospheric circulation

The latitudes of Saturnian jet streams were derived from the wind profile shown in Figure 6 of Del Genio and Barbara (2012). Then the sense of circulation (cyclonic or anti-cyclonic) was inferred from the directionality of each pair of adjacent jets. For example, the pro-grade jet at planetocentric latitude +42 degrees and the retrograde jet at +50 degrees bound a cyclonic flow. These results were converted to planetographic latitudes +48 and +56 degrees. Table 1 (at the end of the manuscript) lists the jet latitudes and indicates whether each interval is cyclonic or anti-cyclonic.

**3. Brightness measurement**

Two hundred HST WFPC images containing the whole disk of Saturn were selected from the MAST archive. These data sample the planet's appearance from near-UV through near-IR wavelengths in a consistent format during a span of 10 years beginning in 1994. When Saturn was imaged repeatedly within a few hours, just one image per filter was selected until the planet rotated by about 120 or 180 degrees. Likewise, when the same planetary longitude was imaged repeatedly over the course of a few days only one image was selected. Thus, the appearance of Saturn was sampled as evenly as possible.

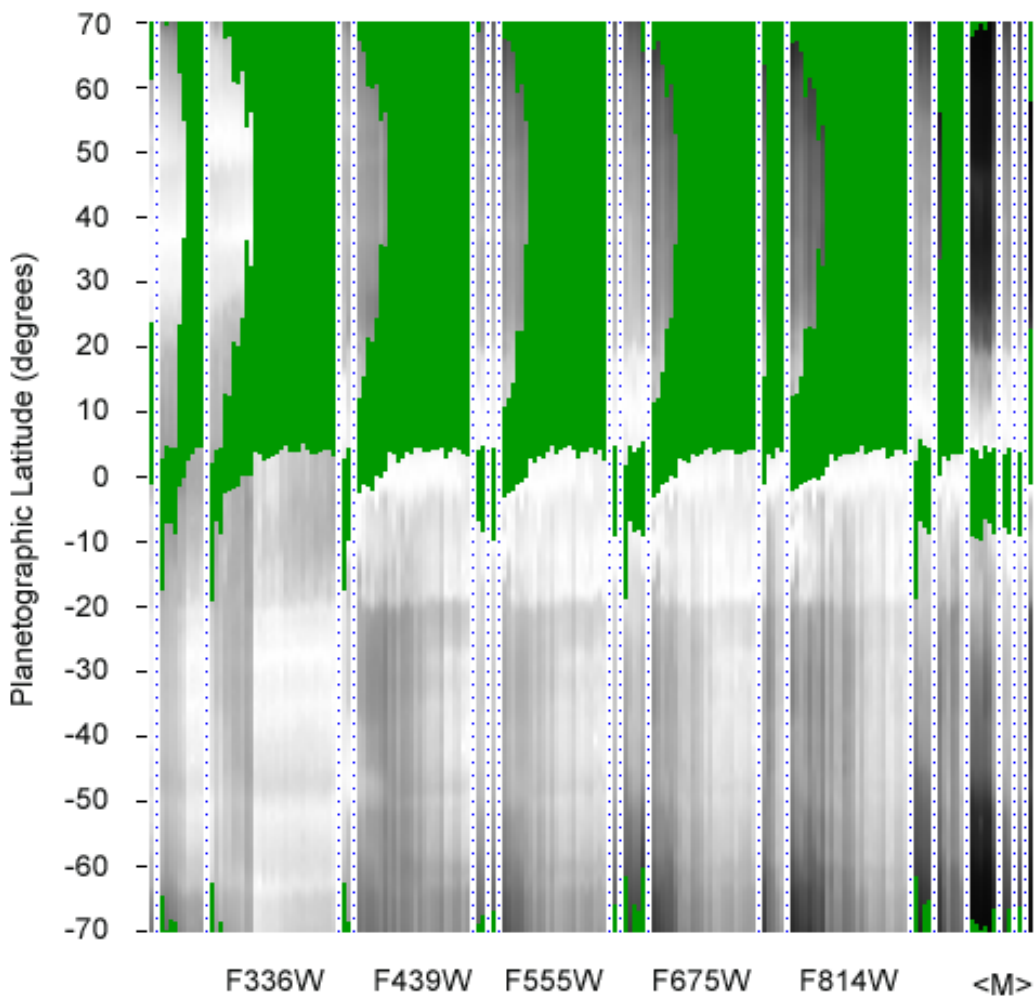

Figure 1. Brightness as a function of the latitude. See the narrative for details.

Brightness was scanned along the meridian, while avoiding the ring system and its shadow. The measures were then normalized to the highest value detected for each image and the pixel locations were mapped to planetographic latitude. The brightness measurements are displayed as a function of latitude and wavelength in Figure 1. Areas in green could not be measured due to the ring system, the shadow of the rings or the inclination of the planet. Results from different filters are separated by vertical dashed lines and wavelength proceeds from left to right. The filters with the greatest number of images are labeled. The methane filters are designed by *<M>*. Within each filter the observations are ordered by time.

## 4. Association between bright bands and cyclonic circulation

The F336W filter emphasizes the contrast between bright and dark bands. The averaged intensity for all those images indicates a strong association between bright banding and cyclonic circulation. Figure 2 demonstrates this over all the observed latitudesand for the seven strongest brightness maxima.

Bright bands at visible and methane wavelengths are less conspicuous, partly because limb darkening is stronger than in the UV. However, such bands are still associated with cyclonic circulation over a wide range of latitudes. Figure3 shows a strong correspondence near the equator. North of the equatorial region bright bands are associated with cyclonic motion at latitudes +48 and +38 degrees, and in the southern hemisphere they are nearly symmetrically placed at -49 and -38 degrees.

The association between bright bands and cyclonic circulation can also be seen when wind profiles are superposed on cylindrical maps of Saturn. The middle and bottom panels of Fig. 5a of Vasavada et al. (2006) are good examples of visible and UV correspondence.

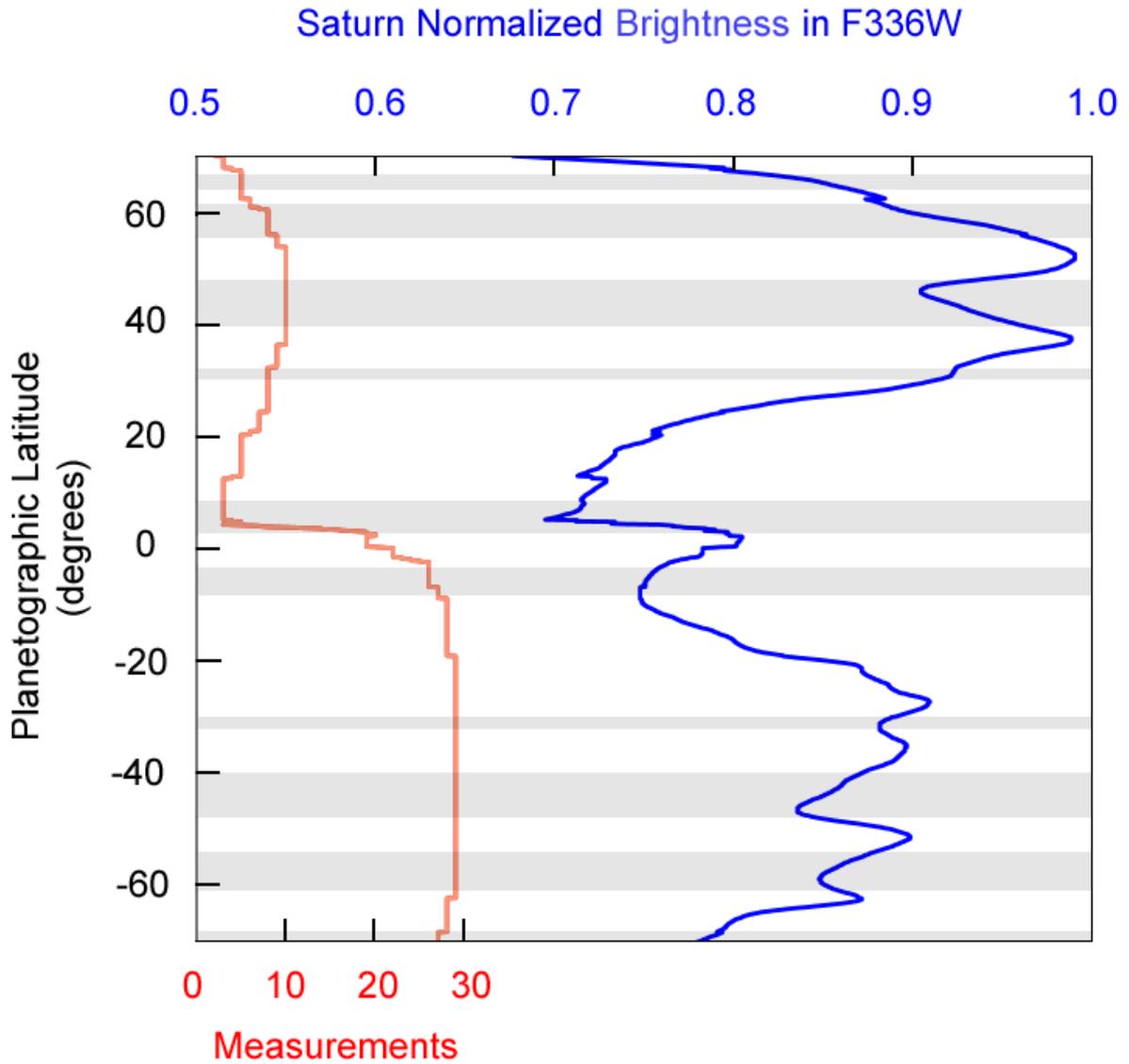

Figure 2. Peaks in UV brightness and cyclonic circulation are strongly associated. The latitudes of anti-cyclonic circulation are shaded grey. The number of images averaged to determine the brightness measurement in each 0.1 degree latitude bin is also indicated.

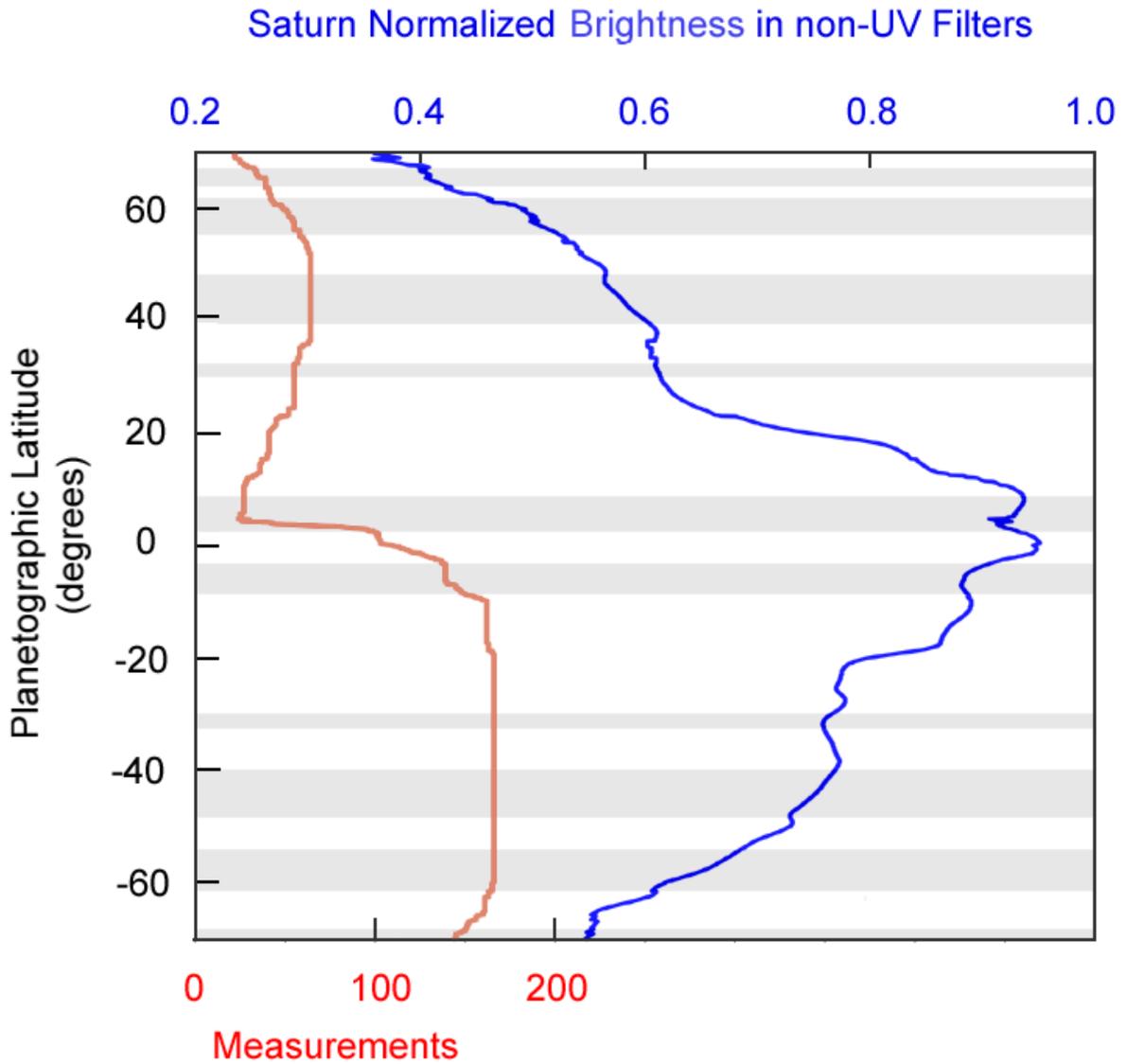

Figure 3. The association between bright bands and cyclonic motion is less apparent in visible light and methane filters due to the reduced local contrast at these wavelengths.

## 5. Comparison with Jupiter

A similar analysis of HST Jupiter images was recently performed by Mallama, Baker and Worley (2014). Two hundred Jovian WFPC images were selected using criteria similar to those described in section 3 of this paper. The presence of large vortices on Jupiter made automated brightness scanning impractical.

Instead, the latitudinal boundaries of bright bands were measured manually. Bright bands were given a weight of 1.0 and uncertain bands were weighted 0.5. Dark bands were weighted 0.0 by default. The average of the values accumulated in each 0.1 degree bin of latitude was referred to as the 'bright band fraction'.

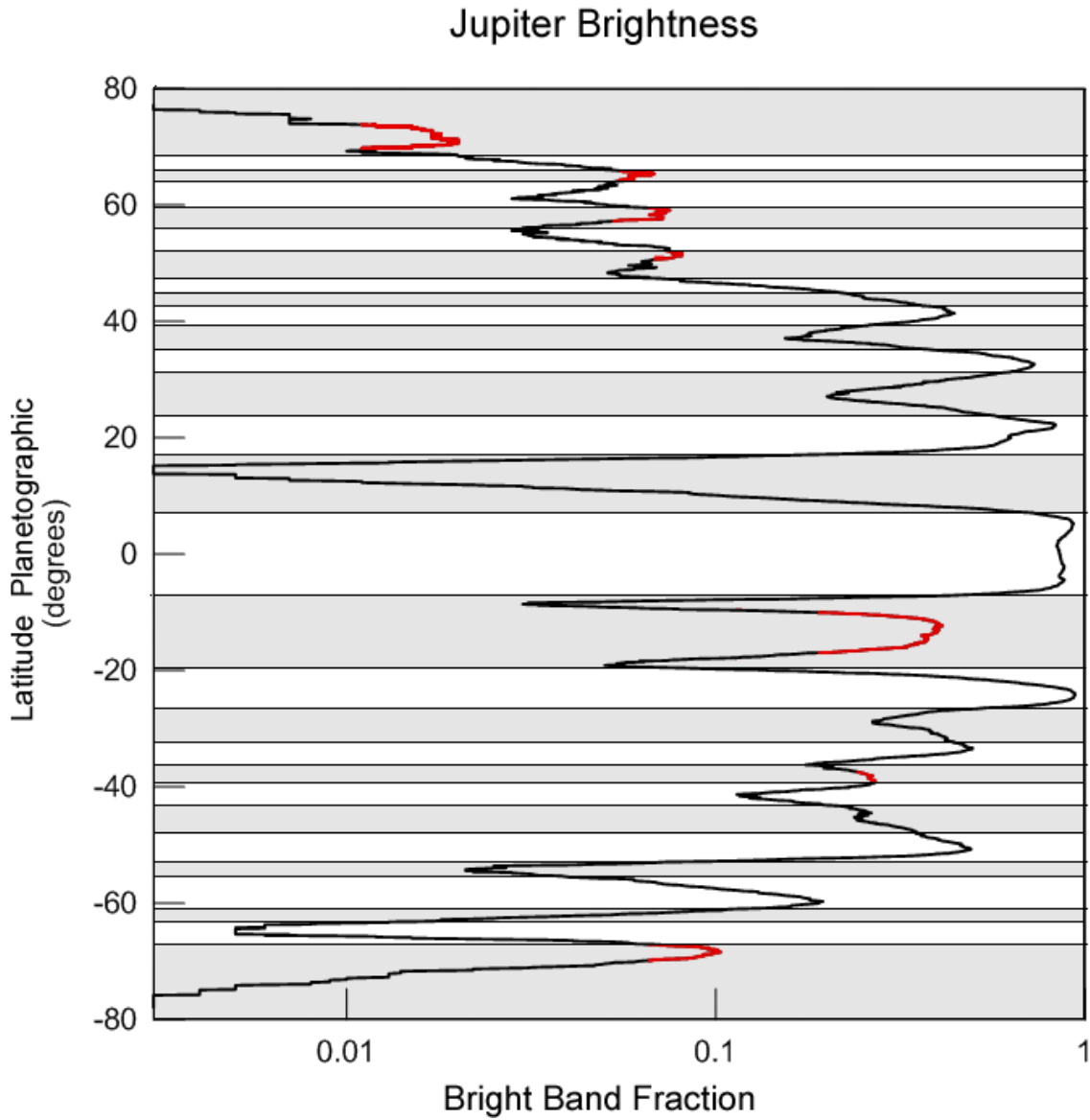

Figure 4. Bright Jovian bands generally align with anti-cyclonic circulation near the equator but not at higher latitudes. Red highlighting indicates bright bands that are in cyclonic regions.

The contrast between bright and dark bands is far stronger than are those on Saturn. Furthermore, the latitudinal boundaries between Jovian bands are practically the same in the F336W filter as for the visible and methane wavelengths. Only UV data shorter than 300 nm appeared anomalous. Therefore the analysis of banding for Jupiter included an average of all data longward of 300 nm.

Bright Jovian bands near the equator were found to be located in regions of anti-cyclonic motion as shown in Figure 4. This is *opposite* to the finding for Saturn. At higher latitudes many bright bands are found in cyclonic regions.

## 6. Definitions of 'belt' and 'zone'

The empirical results reported here highlight the complex relationship between cloud bands and atmospheric circulation. Early attempts to explain band brightness in terms of circulation alone were too simplistic. Atmospheric processes besides circulation must be invoked to account for bright and dark bands.

In fact, the nomenclature once used to distinguish between bright and dark cloud bands no longer relates to brightness. Historically, the bright bands were termed 'zones' while the dark bands were called 'belts' (Rogers, 1995). Now, zones refer to regions of anti-cyclonic circulation while belts are cyclonic regions. Figure 1 of Porco et al. (2003) illustrates this for the case of Jupiter.

## 7. Discussion

This paper presents empirical data that distinguishes the old brightness-related meanings of the terms cloud belt and zone from their new circulation-related meanings. Physical interpretation in the context of cloud layering, morphology and dynamics, however, is outside of its scope. For example, neither the height of the clouds nor the role of aerosols is considered. As such, the parameters of this analysis do not intersect with many of those used in dynamical studies of the giant planets' atmospheres. However, it is important to point out that the that the definitions of the fundamental terms 'belt' and 'zone' have changed.

## 8. Summary

Brightness scans of Saturn imagesobtained by HST indicate a strong association between bright UV bands and cyclonic circulation over a wide range of latitudes. This same association is seen, though not as conspicuously, at non-UV wavelengths. For Jupiter, on the other hand, bright bands near the equator are associated with anti-cyclonic flow. At other Jovian latitudes bright bands are not strongly associated with either direction of atmospheric circulation. The terms 'belt' and 'zone' previouslydistinguished between dark and bright bands, but now they indicate the direction of atmospheric circulation.

*Note:Text files containing the data plotted in the figures of this paper as well as listings of the HST images from which they were derived are available from the author.*

**Acknowledgments**

The author wishes to thank A. Del Genio (Goddard Institute for Space Studies ) for providing helpful information about the cloud bands and atmospheric dynamics of Jupiter and Saturn. M. McIntyre (Cambridge) very kindly reviewed an earlier version of the manuscript. Any errors that may be found in this paper, though, are solely the responsibility of the author. Thanks are also due to the staff of the HST MAST archiveat the Space Telescope Science Institute.

Table 1

Saturnian Jet Latitudes, Circulation and UV Banding

| Jet Latitude (Planetographic) | Relative Direction (By pair) | Cyclonic or Anti- | Bright band? | Dark band? | Comment |
|---|---|---|---|---|---|
| 64.1 | Retrograde | | | | |
| | | Cyclonic | Yes | | Weak band on gradient of strong band |
| 61.3 | Prograde | | | | |
| | | Anti | | Yes | Weak band near cyclonic boundary |
| 55.5 | Retrograde | | | | |
| | | Cyclonic | Yes | | Strong band |
| 47.8 | Prograde | | | | |
| | | Anti | | Yes | Strong band, near cyclonic boundary |
| 39.6 | Retrograde | | | | |
| | | Cyclonic | Yes | | Strong band |
| 31.9 | Prograde | | | | |
| | | Anti | | - | Narrow latitude range |
| 30.3 | Retrograde | | | | |
| | | Cyclonic | - | - | Wide range, mixed banding |
| 8.5 | Prograde | | | | |
| | | Anti | - | - | Few measurements, mixed banding |
| 2.7 | Retrograde | | | | |
| | | Cyclonic | Yes | | Few measurements, strong band |
| 0.0 | Prograde | | | | |
| | | Cyclonic | - | - | Blend with band to north |
| -3.7 | Retrograde | | | | |
| | | Anti | | Yes | Strong band, near cyclonic boundary |
| -8.5 | Prograde | | | | |
| | | Cyclonic | Yes | | Wide range, strong band |
| -30.3 | Retrograde | | | | |
| | | Anti | | Yes | Narrow range, strong band near cyclonic boundary |
| -32.3 | Prograde | | | | |
| | | Cyclonic | Yes | | Strong band |

| | | | |
|---|---|---|---|
| -40.2 | Retrograde | | |
| | Anti | Yes | Strong band, near cyclonic boundary |
| -48.2 | Prograde | | |
| | Cyclonic | Yes | Strong band |
| -54.5 | Retrograde | | |
| | Anti | Yes | Strong band, near cyclonic boundary |
| -61.1 | Prograde | | |
| | Cyclonic | Yes | Strong band |
| -68.7 | Retrograde | | |